# Thermal Conductivity in the Frustrated Two-Leg Spin-Ladder System BiCu$_2$PO$_6$

H Nagasawa[1], T Kawamata[1], K Naruse[1], M Ohno[1], Y Matsuoka[1],

H Sudo[1], Y Hagiya[1], M Fujita[2], T Sasaki[2] and Y Koike[1]

[1]Department of Applied Physics, Tohoku University, Sendai, Japan
[2]Institute for Materials Research, Tohoku University, Sendai, Japan

Email : tkawamata@teion.apph.tohoku.ac.jp

**Abstract**. We have measured temperature and magnetic-field dependences of the thermal conductivity of single crystals of the frustrated two-leg spin-ladder system BiCu$_2$PO$_6$ in magnetic fields up to 14 T. It has been found that the temperature dependence of the thermal conductivity along every principal crystallographic axis shows two peaks in zero field but that the magnitude of the thermal conductivity along the *b*-axis parallel to spin ladders, $\kappa_b$, is significantly larger than those of the thermal conductivity along the *a*-axis, $\kappa_a$, and along the *c*-axis, $\kappa_c$, at high temperatures above 7 K. These results suggest that the thermal conductivity due to spins probably exists only in $\kappa_b$. Furthermore, it has been found that both magnetic-field dependences of $\kappa_a$ and $\kappa_b$ at 3 K show kinks at ~ 7 T and ~ 10 T, where the spin state may change.

## 1. Introduction

In some low-dimensional quantum spin systems, the thermal conductivity has attached great interest, because large contribution of the thermal conductivity due to spins, $\kappa_{spin}$, has been observed [1]. In the two-leg spin-ladder system Sr$_{14}$Cu$_{24}$O$_{41}$, for example, the large contribution of $\kappa_{spin}$ has been observed only along the *c*-axis parallel to the leg direction of spin ladders at high temperatures around 150 K [2,3]. Since the thermal conductivity due to phonons, $\kappa_{phonon}$, has been observed markedly at low temperatures around 20 K along every principal axis, the temperature dependence of the thermal conductivity along the *c*-axis where the magnetic interaction is strong has shown two peaks due to $\kappa_{phonon}$ and $\kappa_{spin}$. Similar two peaks in the temperature dependence of the thermal conductivity have been observed also in several spin-gap systems such as SrCu$_2$(BO$_3$)$_2$ [4,5] and CuGeO$_3$ [6,7]. In these systems, two peaks have been observed in the temperature dependence of the thermal conductivity along every principal axis. Therefore, both two peaks have been concluded to be due to $\kappa_{phonon}$.

Thermal conductivity is a useful probe detecting a change of the spin state in a spin system, because the scattering rate of heat carriers is sensitively affected by the change of the spin state. In fact, anomalies of the thermal conductivity due to the appearance of a new spin state have been observed in several frustrated spin systems [8].

The compound BiCu$_2$PO$_6$ has zigzag two-leg spin-ladders of Cu$^{2+}$ spins with the quantum spin number $S = 1/2$ along the *b*-axis. That is, the leg direction of spin ladders is along the *b*-axis. The

$Cu^{2+}$ spins are frustrated due to both the nearest neighbor exchange interaction along the *b*-axis and the next nearest neighbor exchange interaction along the *b*-axis. The magnitude of the spin gap, $\Delta$, has been estimated from the specific heat, magnetic susceptibility, NMR and magnetization measurements as ~ 35 K [9,10].

In this paper, we have measured the thermal conductivity of $BiCu_2PO_6$ single crystals in magnetic fields to investigate the presence or absence of $\kappa_{spin}$ and also some change of the spin state.

## 2. Experimental

Single crystals of $BiCu_2PO_6$ were grown by the traveling-solvent floating-zone technique. Thermal conductivity measurements were carried out by the conventional steady-state method in magnetic fields up to 14 T at High Field Laboratory for Superconducting Materials, Institute for Materials Research, Tohoku University.

## 3. Results and Discussion

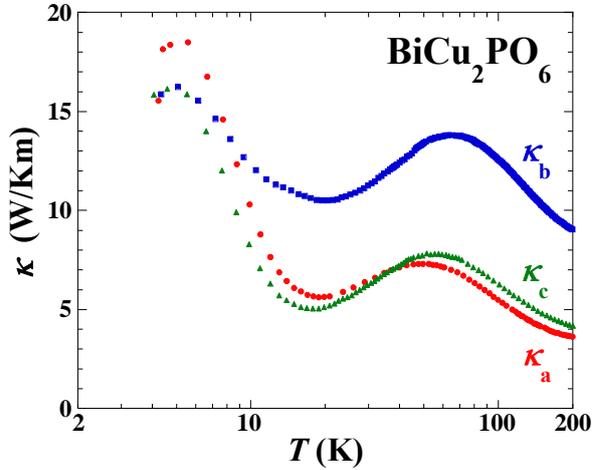

**Figure 1.** Temperature dependence of the thermal conductivity along the *a*-axis, $\kappa_a$, along the *b*-axis, $\kappa_b$, and along the *c*-axis, $\kappa_c$, of $BiCu_2PO_6$ single crystals in zero field.

Figure 1 shows the temperature dependence of the thermal conductivity along the principal crystallographic axes of $BiCu_2PO_6$ single crystals in zero field. It is found that the thermal conductivity along every principal direction shows two peaks, namely, a sharp peak at ~ 5 K and a broad peak at ~ 60 K. Since two peaks are observed in the thermal conductivity along every principal direction, these two peaks are inferred to be due to $\kappa_{phonon}$, as in the case of other spin-gap systems such as $SrCu_2(BO_3)_2$ [4,5]. That is, the broad peak at ~ 60 K is due to the increase of the mean free path of phonons, $l_{phonon}$, with decreasing temperature at high temperatures above ~ 60 K and due to the decrease of $l_{phonon}$ on account of the enhancement of spin fluctuations owing to the frustration, namely, the increase of the phonon-magnon scattering rate with decreasing temperature at low temperatures below ~ 60 K. The sharp peak at ~ 5 K is due to the increase of $l_{phonon}$ on account of the decrease of the phonon-magnon scattering rate owing to the decrease of the number of magnons with decreasing temperature at low temperatures below ~ 20 K where the spin gap is effective. Furthermore, it is found that the magnitude of the thermal conductivity along the *b*-axis parallel to spin ladders, $\kappa_b$, at high temperatures above 7 K is significantly larger than those of the thermal conductivity along the *a*-axis, $\kappa_a$, and along the *c*-axis, $\kappa_c$, perpendicular to spin ladders. Therefore, it is possible that $\kappa_{spin}$ contributes to $\kappa_b$ in a wide region of temperature above 7 K in addition to $\kappa_{phonon}$. There still remains a possibility that the anisotropy of thermal conductivity is ascribed the anisotropy of $\kappa_{phonon}$. To clarify the presence of $\kappa_{spin}$, it may be effective to perform thermal conductivity measurements of $BiCu_{2-x}Zn_xPO_6$, where $\kappa_{spin}$ is expected to be suppressed more strongly than $\kappa_{phonon}$ by nonmagnetic $Zn^{2+}$ ions substituted of $Cu^{2+}$ ions.

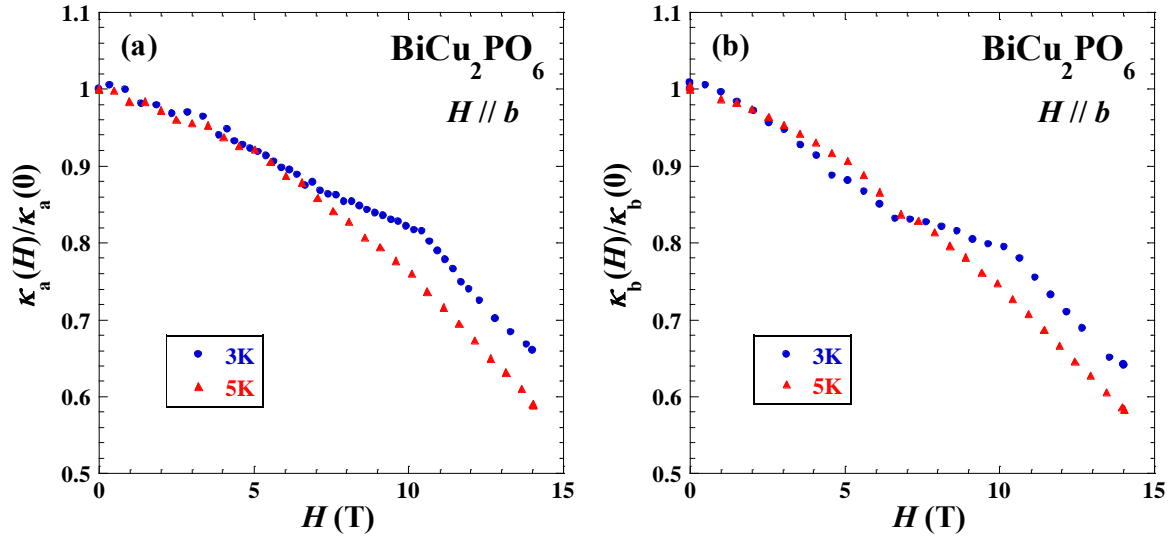

**Figure 2.** Magnetic-field dependence of the thermal conductivity $\kappa_a$ and $\kappa_b$ normalized by the value in zero field, $\kappa_a(H)/\kappa_a(0)$ and $\kappa_b(H)/\kappa_b(0)$, of BiCu$_2$PO$_6$ single crystals in magnetic fields parallel to the $b$-axis up to 14 T at 3 K and 5 K.

Figure 2 shows the magnetic-field dependence of $\kappa_a$ and $\kappa_b$ normalized by the value in zero field, $\kappa_a(H)/\kappa_a(0)$ and $\kappa_b(H)/\kappa_b(0)$, of BiCu$_2$PO$_6$ single crystals in magnetic fields parallel to the $b$-axis. It is found that both $\kappa_a(H)/\kappa_a(0)$ and $\kappa_b(H)/\kappa_b(0)$ are similarly suppressed by the application of magnetic field due to the reduction of the spin gap. At a low temperature of 3 K, it is remarkable that both $\kappa_a(H)/\kappa_a(0)$ and $\kappa_b(H)/\kappa_b(0)$ exhibit kinks at ~ 7 T and ~ 10 T. Since the kinks are observed in both $\kappa_a$ and $\kappa_b$, they will be due to the change of $\kappa_{phonon}$ via the phonon-magnon scattering caused by the change of the spin state. However, there has been no report on the anomaly at ~ 7 T and ~ 10 T at low temperatures, through field-induced magnetically ordered states have been reported to appear in magnetic fields as high as ~ 20 T and ~ 35 T [11]. Very recent inelastic neutron scattering experiments in magnetic fields parallel to the $a$-axis have revealed that there is a level crossing of magnon branches, which are split by the anisotropic interaction in zero field, around 11T [12]. Therefore, the kinks of $\kappa_a(H)/\kappa_a(0)$ and $\kappa_b(H)/\kappa_b(0)$ in magnetic fields parallel to the $b$-axis might be due to such a level crossing. In any case, this result suggests to perform detailed experiments in magnetic fields below ~ 10 T at low temperatures to clarify the possible change of the spin state.

## 4. Conclusions

We have measured temperature and magnetic-field dependences of the thermal conductivity of BiCu$_2$PO$_6$ single crystals in magnetic fields up to 14 T. In zero field, it has found that the temperature dependence of the thermal conductivity along every principal crystallographic axis shows two peaks. These peaks have been explained as being due to the temperature dependence of $\kappa_{phonon}$, namely, the temperature dependence of the phonon-magnon scattering rate, taking into account the spin fluctuations owing to the frustration and the spin gap. It has been found that it is possible that $\kappa_{spin}$ contributes to $\kappa_b$ in the leg direction of spin ladders, because $\kappa_b$ is significantly larger than $\kappa_a$ and $\kappa_c$ at high temperatures above 7 K. To clarify the contribution of $\kappa_{spin}$, thermal conductivity measurements of Zn-substituted BiCu$_{2-x}$Zn$_x$PO$_6$ may be effective. It has also been found that there are two kinks in the magnetic-field dependence of $\kappa_a(H)/\kappa_a(0)$ and also $\kappa_b(H)/\kappa_b(0)$ at ~ 7 T and ~ 10 T at a low temperature of 3 K. These kinks suggest some change of the spin state or a level crossing of magnon branches. To clarify them, other detailed experiments are desired in magnetic fields below ~ 10 T at low temperatures.